\documentclass[12pt,notitlepage,a4paper]{article}

\usepackage{epsfig}
\usepackage{color,graphicx}
\usepackage{cite}
\usepackage{amssymb,amsmath}
\usepackage{epstopdf}

\newcommand{\be}{\begin{equation}}
\newcommand{\ee}{\end{equation}}
\newcommand{\bea}{\begin{eqnarray}}
\newcommand{\eea}{\end{eqnarray}}

\begin{document}

\begin{center}  

\vskip 2cm 

\centerline{\Large {\bf Quantum Hall Effect in a Holographic Model}}
\vskip 1cm

\renewcommand{\thefootnote}{\fnsymbol{footnote}}

\centerline{Oren Bergman,${}^1$\footnote{bergman@physics.technion.ac.il} 
Niko Jokela,${}^{1,2}$\footnote{najokela@physics.technion.ac.il} 
Gilad Lifschytz,${}^2$\footnote{giladl@research.haifa.ac.il} 
and Matthew Lippert${}^3$\footnote{mlippert@physics.uoc.gr}}

\vskip .5cm
${}^1${\small \sl Department of Physics} \\
{\small \sl Technion, Haifa 32000, Israel} 

\vskip 0.2cm
${}^2${\small \sl Department of Mathematics and Physics} \\
{\small \sl University of Haifa at Oranim, Tivon 36006, Israel} 

\vskip 0.2cm
${}^3${\small \sl Crete Center for Theoretical Physics} \\
{\small \sl Department of Physics} \\
{\small \sl University of Crete,  71003 Heraklion, Greece}

\end{center}

\vskip 0.3 cm

\setcounter{footnote}{0}
\renewcommand{\thefootnote}{\arabic{footnote}}

\begin{abstract}
\noindent We consider a holographic description of a system of strongly coupled fermions in 
$2+1$ dimensions based on a D7-brane probe in the background of D3-branes,
and construct stable embeddings by turning on worldvolume fluxes.
We study the system at finite temperature and charge density,
and in the presence of a background magnetic field.
We show that Minkowski-like embeddings that 
terminate above the horizon describe a family of quantum Hall states with 
filling fractions that are parameterized by a single discrete parameter.
The quantization of the Hall conductivity is a direct consequence of the
topological quantization of the fluxes.
When the magnetic field is varied relative to the charge density away from these discrete 
filling fractions, the embeddings deform continuously into black-hole-like embeddings
that enter the horizon and that describe metallic states.
We also study the thermodynamics of this system and 
show that there is a first order phase transition at a critical temperature
from the quantum Hall state to the metallic state.
\end{abstract}

\newpage

\section{Introduction}

Fermions at strong coupling exhibit many interesting phenomena that are qualitatively different 
from those of weakly 
coupled fermions and are therefore very difficult to describe theoretically.
The fractional quantum Hall effect (FQHE) is an example of such a phenomenon in 2+1 dimensions.

Electrons moving in effectively two spatial dimensions and subject to a strong magnetic field
exhibit plateaus in the transverse (Hall) conductivity as the magnetic field is varied.
These plateaus occur around certain rational values of the Landau level filling fraction 
$\nu$, defined as the ratio of the charge density to the magnetic field.
At these values, the longitudinal conductivity vanishes, indicating the formation of a gapped state.
Away from these values of $\nu$, the state is ungapped, the longitudinal conductivity is non-zero, and the Hall
conductivity varies.
The plateaus at integer values of $\nu$ are well understood in terms of the physics 
of free electrons in a background magnetic field (the Landau problem).
More precisely, the existence of a gapped state at integer filling fractions is explained
in this way, but the non-vanishing width of the plateau around these values requires impurities,
which give rise to localized states in the gap.

In contrast, the plateaus at non-integer filling fractions (the FQHE) cannot be explained 
in this way, and their existence seems to depend crucially on strong interaction dynamics.
The fractional quantum Hall fluid, as this strongly coupled state is known, really corresponds to a new 
state of electron matter.
A phenomenological model, based on a variational wave-function, 
describing this state for $\nu=1/(2k+1)$ was proposed by Laughlin \cite{Laughlin:1983fy}.
The Laughlin state is gapped, and the lowest lying excitations are
quasiparticles carrying fractional charge $\nu e$.
This remarkable property was observed experimentally \cite{quasiparticles}, 
confirming the validity of Laughlin's proposal.
However, many questions remain unanswered;  for example, we still seek a complete understanding 
of the allowed
filling fractions and the nature of the transitions between the different plateaus.
It is likely that these questions require a better understanding of the microscopic physics.

Our goal is to exhibit a quantum Hall effect in a strongly coupled system of fermions,
namely one in which the weak coupling explanation via Landau levels does not work. 
Gauge/gravity duality has emerged as a powerful new tool for analyzing a class
of strongly interacting systems that are described (at weak coupling) by large $N$ gauge field theories.
Of particular interest are models with only light fermions in the fundamental representation, 
which could, in principle, provide microscopic toy models for fermionic matter at strong coupling.
These models are based on brane configurations with two sets of D-branes, such that there are six or eight
directions with mixed boundary conditions on the worldsheet ($\#ND=6,8$).
In these cases one generally works in the probe approximation, where the number of branes
in one set is much greater than the other. The first set of branes then provides the near-horizon 
gravitational background,
and the other branes are treated as probes in this background.
The much-studied Sakai-Sugimoto D4-D8 model is an example of such a system, which describes at low energy
a four-dimensional QCD-like theory with $N_4\rightarrow\infty$ colors  and $N_8\ll N_4$ flavors of massless quarks \cite{SS}.
The analogous system in three dimensions consists of D3-branes and D7-branes that have $\#ND=6$ and therefore
share three spacetime dimensions.
As in the D4-D8 model, supersymmetry is completely broken, and the light flavor degrees of freedom 
are purely fermionic.
This system, therefore, has the potential to provide a gravitational dual of a (strongly coupled) quantum Hall fluid.
However, this does not quite work in the simplest setup, where the D7-brane wraps an $S^4$ inside the $S^5$
of the near-horizon D3-brane background, due an unstable ``slipping" mode \cite{Rey:2008zz}. 
This is simply a manifestation of the repulsive interaction between the two kinds of branes 
in flat space. Nevertheless, some aspects of the quantum Hall effect have been modeled in this system
by considering the D7-brane embedding in the asymptotically flat, rather than near-horizon, 
D3-brane background and varying the mass of the fermion rather than the magnetic field, 
which gives a toy model for a plateau transition \cite{perkraus}.
This D3-D7 setup has also been used to study some aspects of large $N$ three-dimensional QCD
in \cite{Hong:2010sb}.

We will consider a slightly different D7-brane geometry, in which the D7-brane wraps an $S^2\times S^2$
inside the $S^5$.
Following the idea of \cite{Myers:2008me}, we will show that by turning on appropriate worldvolume 
fluxes on the $S^2$'s the embeddings can be stabilized. 
We find two families of embeddings at finite temperature: ``black hole embeddings" which
enter the horizon and ``Minkowski embeddings" which avoid it. 
We analyze the properties of both types at finite charge density and background magnetic 
and electric fields.
The black hole embeddings correspond to metallic states with a non-vanishing longitudinal
conductivity and an unquantized transverse conductivity.

The Minkowski embeddings on the other hand,  which exist only when the ratio of charge density to 
magnetic field takes particular quantized values, describe gapped states with a vanishing 
longitudinal conductivity and a transverse conductivity proportional to the ratio of the charge density to the magnetic field. 
We should stress that these are not the quantized filling fractions observed in experiments;
however, this model achieves our goal of exhibiting a quantum Hall effect in a strongly
coupled system.
Furthermore, in the holographic description the quantization is topological since it
originates from the Dirac quantization of the magnetic fluxes on the $S^2$'s.
In particular, the transverse conductivity in the Minkowski embeddings is independent of 
the temperature.
To some extent, this is what is seen in the real quantum Hall effect:
the quantization of the transverse conductivity is robust against small environmental changes.
This is understood in terms of topology in the integer case \cite{Thouless}.

Other holographic models for the quantum Hall effect have appeared in 
\cite{KeskiVakkuri:2008eb,Fujita:2009kw,Hikida:2009tp}.
In fact, the realization of the quantum Hall effect using branes in string theory has a long history, beginning with \cite{Brodie:2000yz,Hellerman:2001yv,Bergman:2001qg}.

\medskip

The rest of the paper is organized as follows: 
In Section \ref{sec:setup} we discuss the D7-brane embeddings and their stability.
In Section \ref{sec:qhstate} we focus on the Minkowski embeddings and show that they 
describe quantum Hall states with discrete filling fractions.
In Section \ref{sec:metallic} we analyze the black hole embeddings, which correspond to the metallic states.
In Section \ref{sec:thermo} we study the thermodynamics of the system.
Finally, we end the paper with a discussion section, in which
we raise a number of issues related to our model, and
suggest how to make contact with other well-known properties of the 
fractional quantum Hall effect, including the fractionally charged quasiparticles,
edge states, and the role of impurities.
The details of some of the calculations appear in the appendices.

\section{The D3-D7' system}\label{sec:setup}

A simple brane configuration that realizes charged fermions, and no charged bosons, in 2+1 dimensions 
at low energy consists of a D3-brane and a D7-brane arranged as follows:
\be
\label{flatD3D7}
\begin{tabular}{ccccccccccc}
& 0 & 1 & 2 & 3 & 4 & 5 & 6 & 7 & 8 & 9 \\
D3 & $\bullet$ & $\bullet$ & $\bullet$ & $\bullet$ & & & & &  \\
D7 & $\bullet$ & $\bullet$ & $\bullet$ &  & $\bullet$ & $\bullet$ & $\bullet$
&  $\bullet$ & $\bullet$ &  \\
\end{tabular}
\ee
The D3-D7 open strings have $\#ND=6$, which means that the NS sector is massive and only
the R sector contains massless states. 
In this case each D3-D7 pair gives a complex, two-component spinor
whose mass is determined by the separation in the common transverse direction $x^9$.
This configuration is non-supersymmetric and unstable  since the branes are also repelled from one another
in this direction.
The dual bulk description is obtained by taking a large number of D3-branes
and a finite number of D7-branes, \emph{i.e.}, $N_7\ll N_3$.
In this case the D7-branes can be treated as probes in the near-horizon 
$AdS_5\times S^5$ background of the D3-branes. 
The background describes the four-dimensional gauge field dynamics, and the D7-brane embedding
in that background captures the three-dimensional physics of the fermions.
Starting with the above flat brane configuration and taking the near-horizon limit, the D7-branes
wrap an $S^4 \subset S^5$, and are extended along $AdS_4 \subset AdS_5$.
The instability now appears as a tachyonic mode,
which violates the Breitenlohner-Friedman (BF) bound for $AdS_4$, for the D7-branes to ``slip off" the $S^5$.

\subsection{$S^2\times S^2$ embedding}

Let us consider a slightly different embedding, in which the D7-brane wraps an $S^2\times S^2\subset S^5$.
(We focus on the case of a single D7-brane.)
We begin with the near-horizon background of the non-extremal D3-brane:
\begin{eqnarray}
\label{D3metric}
 L^{-2} ds_{10}^2 &=& r^2 \left(-h(r)dt^{2}+dx^2+dy^2+dz^2\right)+
 r^{-2} \left(\frac{dr^2}{h(r)}+r^2 d\Omega_5^2\right) \\
\label{RR_5-form}
F_5 &=& 4L^4\left(r^3 dt\wedge dx\wedge dy\wedge dz\wedge dr 
+  d\Omega_5 \right) \,,
\end{eqnarray}
where $h(r)=1-r_T^4/r^4$ and $L^2=\sqrt{4\pi g_{s} N_3}\, \alpha'$. 
For convenience, we work in dimensionless coordinates, {\em e.g.}, $r=r_{phys}/L$.
In general we will use lower case latin letters for dimensionless quantities,
and when needed we will use upper case letters for their physical counterparts.
This background is dual to ${\cal N}=4$ SYM theory at a temperature $T = r_T/(\pi L)$.
We parameterize the five-sphere as an $S^2\times S^2$ fibered over an interval:
\bea
 d\Omega_5^2 &=& d\psi^2 + 
 \cos^2\psi (d\Omega_2^{(1)})^2 + \sin^2\psi (d\Omega_2^{(2)})^2 \nonumber \\
 (d\Omega_2^{(i)})^2 &=& d\theta_i^2 + \sin^2\theta_i d\phi_i \,,
\eea
where $0\leq \psi \leq \pi/2$, $0\leq \theta_i \leq \pi$, and $0\leq \phi_i < 2\pi$.
As $\psi$ varies, the sizes of the two $S^2$'s change. At $\psi=0$ one of the $S^2$'s shrinks to zero
size, and at $\psi=\pi/2$ the other $S^2$ shrinks. The $S^2\times S^2$ at $\psi=\pi/4$ is the ``equator".
It will also be useful to have an explicit expression for the RR 4-form potential.
In a partially fixed gauge we can take the 4-form to be
\be
\label{RR_4-form}
C_4 = L^4 \left(r^4\, dt\wedge dx\wedge dy\wedge dz
+ {1\over 2} c(\psi)\, d\Omega_2^{(1)}\wedge d\Omega_2^{(2)}\right) \,,
\ee
where $c(\psi)\equiv (8\pi^2 L^4)^{-1}\int_{S^2\times S^2} C_4$. Up to an additive 
constant corresponding to the residual gauge freedom, this is given by 
\be
\label{4-form_flux}
c(\psi) = \psi - {1\over 4}\sin 4\psi + \rm{const} \,.
\ee
We will fix this constant later.

The D7-brane wraps the two $S^2$'s and is extended along $t,x,y$, and $r$.
Its embedding is then described by the two scalar fields $\psi(r)$ and $z(r)$.
The induced metric on the D7-brane is given by
\bea
 L^{-2} ds^2_{D7} & = & r^2 \left(-h(r)dt^2+dx^2+dy^2\right)
 +\left(\frac{1}{r^2 h(r)} + r^2 z'(r)^2 + \psi'(r)^2\right)dr^2 \nonumber \\
 & & + \cos^2\psi (d\Omega_2^{(1)})^2 + \sin^2\psi (d\Omega_2^{(2)})^2 \,.
\eea
The worldvolume DBI action is then 
\be
 S_{DBI}  =  
 -4{\cal N} \int dr\, r^2 \cos^2\psi \sin^2\psi
  \sqrt{1+r^4h(r) z'^2+r^2 h(r)\psi'^2} \,,
\ee
where 
\be
{\cal N} \equiv 4\pi^2 L^5 T_7 V_{2,1} \,.
\ee
We have absorbed the volume of spacetime $V_{2,1}$ into the normalization factor,
since our Lagrangian densities will never have a spacetime dependence.
This gives the following coupled equations for $z(r)$ and $\psi(r)$:
\bea
\partial_r\left({r^4 g(r)} z'(r)\right) &=& 0 \\
\label{psieom1}
\partial_r\left({r^2 g(r)}\, \psi'(r)\right) & = & 
{32 r^4 h(r) \over g(r)} \cos^3\psi \sin^3\psi 
\left(\cos^2\psi - \sin^2\psi\right)  \,,
\eea
where we have defined
\be
g(r) \equiv {4r^2h(r) \cos^2\psi \sin^2\psi\over\sqrt{1 + r^4 h(r) z'^2+ r^2 h(r)\psi'^2}} \,.
\ee
The equation for $z(r)$ can be integrated once to obtain 
\be
\label{zeom}
r^4 g(r) z'(r) = c_z \,.
\ee
At this point the constant of integration $c_z$ is arbitrary, but later we will show that it is
fixed by regularity conditions on the solutions.

The equation for $\psi(r)$ has three constant solutions: $\psi =0, \pi/2$, and $\pi/4$.
The first two are trivial since the D7-brane has a vanishing size.
In the $\psi=\pi/4$ case the D7-brane wraps the ``equatorial" $S^2\times S^2$.
As was the case for the $S^4$ embedding, this embedding is unstable to ``slipping"
towards one of the trivial solutions.
This can be shown by an analysis of the fluctuations, which reveals a tachyonic mode
that violates the BF bound.
Alternatively, one can look at the asymptotic form of non-constant solutions
and read off the dimension of the corresponding operator, in which case
the instability shows up as a non-zero imaginary part.
This method is somewhat simpler, especially when we turn on background gauge fields later.
Plugging in the ansatz
\be
\label{ansatz}
\psi(r) \sim {\pi\over 4} + Ar^\Delta
\ee
into the large $r$ asymptotic form of (\ref{psieom1}) gives
\be
\Delta(\Delta + 3) = -8 \,,
\ee
which does not have a real solution. 
The lowest mode of the field $\psi$ is a tachyon with mass-squared = $-8L^{-2}$,
which violates the BF bound for $AdS_4$ of $-(9/4)L^{-2}$.
This aspect of the $S^2\times S^2$ embedding is no different from the $S^4$ embedding.

\subsection{Flat space brane configuration and the fermion mass}

\begin{figure}[ht]
\centerline{\epsfig{file=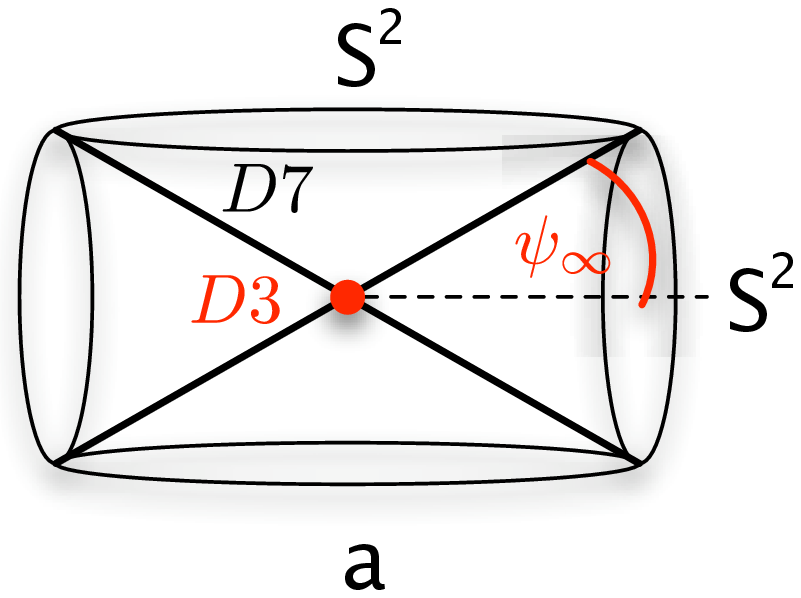,height=3cm}\hspace{1cm}\epsfig{file=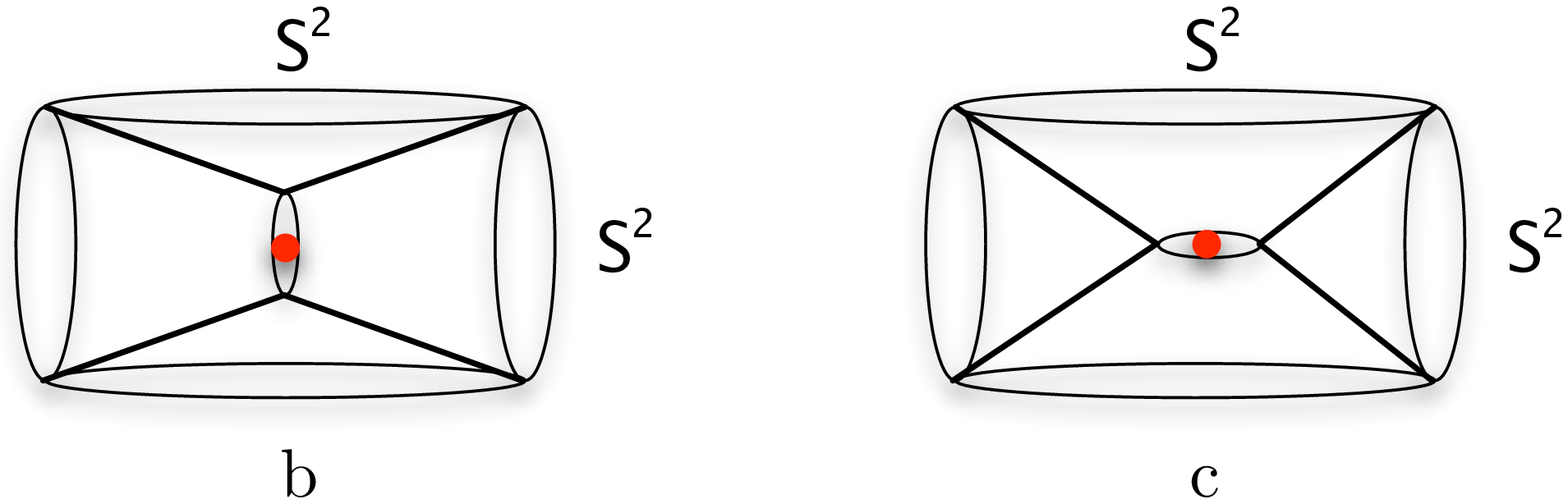,height=3cm}}
\caption{Brane configuration for (a) massless fermions and (b), (c) massive fermions.}
\label{brane_config}
\end{figure}

Before moving on to stabilize the embedding, let us address an obvious question:
what is the brane configuration in flat space that leads to this embedding in the decoupling limit? 
We need to know this, for example, in order to identify the fermion mass parameter 
in terms of the embedding fields.
For the $S^4$ embedding this was just the flat D3-D7 configuration (\ref{flatD3D7}).
In that case the D7-brane spanned an $R^5$ subspace of the $R^6$ transverse to the D3-brane.
For an $S^2\times S^2$ embedding in the decoupling limit, we have to start with
a D7-brane that spans a cone over $S^2\times S^2$ in $R^6$, with the D3-branes
at the (singular) origin (see Fig.~\ref{brane_config}a).\footnote{This is a five-dimensional version of the conifold.}
There is a one-parameter family of such geometries parameterized by the relative size
of the two $S^2$'s in the base of the cone, which we identify with the angle $0\leq\psi\leq \pi/2$.
Only the symmetric $\psi=\pi/4$ configuration is a non-trivial solution, albeit an unstable one, 
since otherwise there is a non-zero force between different parts of the D7-brane.
However, we will keep $\psi$ arbitrary for now.
This configuration gives massless fermions in 2+1 dimensions.
There are two possible deformations that give the fermions a mass, corresponding to blowing up either of the
$S^2$'s at the origin (see Fig.~\ref{brane_config}b,c). 
These are also the instabilities of the configuration in the $\psi=\pi/4$ case.
In these configurations the angle $\psi$ varies with the distance to the D3-branes
and asymptotes to its value in the massless configuration, which we will denote $\psi_\infty$.
The fermion mass is given by the minimal distance between 
the D3-brane and the D7-brane, which is easily computed using trigonometry (recall that we are using
dimensionless coordinates):
\be
{2\pi\alpha'\over L} M = {r\sin(\psi(r)-\psi_\infty)\over \cos\psi_\infty} \;\; \mbox{or} \;\;
 {r\sin(\psi_\infty - \psi(r))\over \sin\psi_\infty} \,,
\ee
for the first and the second deformation, respectively.



\subsection{Stable embeddings}

In order to stabilize the embedding, we will employ the method proposed in \cite{Myers:2008me},
by turning on a background worldvolume gauge field on the D7-brane.
In our case we turn on magnetic fields on the  two $S^2$'s:
\be
2\pi\alpha' F = {L^2\over 2} \left(f_1 d\Omega_2^{(1)} + f_2 d\Omega_2^{(2)}\right)\,.
\ee
The numbers $f_1,f_2$ are quantized as
\be
f_i = {2\pi\alpha'\over L^2}\, n_i \,,
\ee
where $n_i$ are integers.
The D7-brane DBI action is now:
\be
 S_{DBI} =  -{\cal N}\int dr\, r^2\sqrt{\left(4\cos^4\psi + {f_1^2} \right)
 \left(4\sin^4\psi + {f_2^2} \right) \left(1+r^4 h z'^2+r^2 h \psi'^2\right)} \;.
\ee
In addition, there is now a non-zero CS term:
\be
S_{CS} = -{(2\pi\alpha')^2T_7\over 2} \int P[C_4]\wedge F \wedge F \,,
\ee
where $P[C_4]$ is the pullback to the D7-brane worldvolume of the background RR 4-form potential.
It is worth noting at this point that the CS action, as written above, is not gauge invariant
under gauge transformations of $C_4$. The action transforms by surface terms,
which in principle should be cancelled by the addition of a boundary action.
Usually one ignores these in infinite volume, but if some of the fields
are non-vanishing at infinity, they may be relevant.
For the present discussion we can ignore the boundary terms, but they will be important 
in the next subsection.
Pulling back the RR 4-form potential (\ref{RR_4-form}) gives
\be 
\label{CS1}
 S_{CS} = - {\cal N} f_1 f_2 \int dr\, r^4 z'(r) \,.
\ee
The new equations of motion for $z(r)$ and $\psi(r)$ are:
\bea
\label{zeom2}
g(r)z' & = & \frac{c_z}{r^4}-f_1 f_2 \\
\label{psieom2}
\partial_r\left({r^2 g(r)}\, \psi'(r)\right) & = & {8 r^4 h(r) \over g(r)} \cos\psi \sin\psi \times \nonumber\\
&& \qquad \times \left[(f_1^2 + 4\cos^4\psi)\sin^2\psi - (f_2^2 +4\sin^4\psi)\cos^2\psi\right]  \,,
\eea
where the definition of $g(r)$ has been changed to incorporate the fluxes $f_1$ and $f_2$:
\be
g(r) \equiv {r^2 h(r) \sqrt{(f_1^2 + 4\cos^4\psi)(f_2^2 + 4\sin^4\psi)}
\over\sqrt{1 + r^4 h(r) z'^2 + r^2 h(r) \psi'^2}} \,.
\ee

We would like to determine under what conditions on $f_1$ and $f_2$ there exist stable embeddings
with the asymptotic behavior 
\be
\psi(r) \sim \psi_{\infty}+mr^{\Delta_{+}}-c_{\psi}r^{\Delta_{-}}.
\ee
The solution of the $\psi$ equation of motion to leading order at large $r$ gives
either $\psi_\infty = 0$ or $\pi/2$ for any $f_1,f_2$, or
\be
\label{const_solution}
(f_1^2 + 4\cos^4\psi_\infty)\sin^2\psi_\infty = (f_2^2 + 4\sin^4\psi_\infty)\cos^2\psi_\infty \,.
\ee
We will concentrate on the latter type of solution.
Looking at the subleading terms then gives
(see Appendix \ref{app:stability} for  details)
\be
\Delta_\pm = -\frac{3}{2}\pm \frac{1}{2}\sqrt{9+16\frac{f_1^2+16\cos^6\psi_\infty-12\cos^4\psi_\infty}{f_1^2+4\cos^6\psi_\infty}} \ .
\label{deltapm}
\ee
The embedding is stable if $\Delta_{\pm}$ is real.
The operator dual to $\psi(r)-\psi_\infty$ in the field theory is the fermion bi-linear.
If $\Delta_+=-1$ and $\Delta_-=-2$, we can identify the parameter $m$ with the mass of the fermion 
(up to a multiplicative constant) and the parameter $c_\psi$ with the fermion bi-linear
condensate. 
However, more general embeddings are possible.
We will require that $\Delta_\pm<0$, since otherwise the D7-brane intersects itself (an infinite number of times).

On general grounds, there can be two types of embeddings,
depending on the form of the solution to the $\psi(r)$ equation (\ref{psieom2}).
The ``black hole" (BH) embeddings correspond to solutions in which the D7-brane 
crosses the horizon at $r=r_T$,
and the ``Minkowski" (MN) embeddings correspond to solutions in which the D7-brane
terminates smoothly at some $r=r_0>r_T$,
which means that $\psi(r_0) = 0$ or $\pi/2$, corresponding to one or the other $S^2$ shrinking.
We will examine the two types of embeddings and determine their physical meaning in Sections \ref{sec:qhstate} and \ref{sec:metallic}. 
But, first we would like to add two more ingredients: charge and magnetic field.

\subsection{Finite charge density and magnetic field}

Boundary charge currents and electromagnetic fields are both encoded in the D7-brane worldvolume 
gauge field. Strictly speaking, the currents are global and the fields are not dynamical, but we can
still study the effect of a background electromagnetic field on the currents.
Here we will consider a background magnetic field described by a spatial component of the D7-brane gauge field 
and a non-zero charge density described by the time component:
\be
A_y = {L\over 2\pi\alpha'} xb \;\; , \;\; A_0 = {L\over 2\pi\alpha'} a_0(r) \,.
\ee
In keeping with our convention, the quantities $b$ and $a_0$, as well as $x$, are dimensionless.
Thus, the physical magnetic field is given by $B=b/(2\pi\alpha')$.
The DBI action is now
\bea
\label{DBI3}
 S_{DBI} & = & - {\cal N} \int dr\, r^2\sqrt{\left(4\cos^4\psi + f_1^2 \right)
 \left(4\sin^4\psi + f_2^2 \right)}\times \nonumber \\ 
 & & \qquad\qquad \times \sqrt{\left(1+ r^4 h(r) z'^2+ r^2 h(r)\psi'^2 - {a_0'}^2\right)\left(1+\frac{b^2}{r^4}\right)} \,,
\eea
and the CS action is
\bea
\label{CS2}
S_{CS} &=& -{(2\pi\alpha')^2T_7\over 2} \int P[C_4]\wedge F \wedge F \nonumber\\
&=&  -{\cal N}f_1 f_2 \int dr\, r^4 z'(r)
+ 2{\cal N} \int dr\, c(r) b a_0'(r) \,,
\eea
where $c(r)\equiv c(\psi(r))$.
In other words, $c(r)$ measures the flux of the RR 4-form potential on the $S^2\times S^2$
that the D7-brane occupies at radial position $r$.
Recall, however,  that the CS action is only gauge invariant up to surface terms, and one needs
to add boundary terms to the action, such that their gauge variation cancels the surface terms.
In particular, in the partially fixed gauge of (\ref{RR_4-form}) there is a residual gauge symmetry
that shifts $c(r)$ by a constant. 
We must therefore add the following boundary term to the action:\footnote{This boundary term
can also be obtained by starting with the alternative form of the CS action
$\int P[F_{5}]\wedge A \wedge F$ and integrating by parts in the radial coordinate.
In this form the CS action is invariant under the RR gauge transformation, 
but not under the gauge transformation of the worldvolume gauge field,
and therefore also requires the addition of (other) boundary terms.
See for example \cite{Bergman:2008qv} for a related discussion in the Sakai-Sugimoto model.}
\be
S_{boundary} = - 2{\cal N} c(r) b a_0(r)\Big|_{r_{min}}^\infty \,.
\ee
Let us finally fix the gauge completely by requiring $c(r\rightarrow \infty) = 0$.
This means that
\be
\label{4-form_flux_2}
c(r) = \psi(r) - {1\over 4}\sin\left(4\psi(r)\right) - \psi_\infty + {1\over 4}\sin(4\psi_\infty)  
\ee
and corresponds to the amount of (gauge invariant) 5-form flux that permeates the D7-brane between $r$
and infinity.
The boundary term is then just
\be
S_{boundary} =  2{\cal N} c(r_{min}) b a_0(r_{min}) \,.
\label{boundary_action}
\ee
The quantity $c(r_{min})$ is the total amount of 5-form flux captured by the D7-brane.
The boundary term does not contribute to the equations of motion, but it will contribute to the on-shell action,
and therefore to the thermodynamic potentials.

The equations for $z(r)$ and $\psi(r)$ are now 
\be
\label{zeom3}
g(r)\left(1 + {b^2\over r^4}\right) z'(r) = \frac{c_z}{r^4}-f_1 f_2 
\ee
\bea
\label{psieom3}
 & & \partial_r\left( r^2 g(r)\left(1 + {b^2\over r^4}\right) \psi'(r)  \right) = -16\cos^2\psi\sin^2\psi b a'_0
 \nonumber \\
       &&   \mbox{}  + \frac{8h(r)r^4}{g(r)} \cos\psi\sin\psi \left[(f_1^2 + 4\cos^4\psi)\sin^2\psi 
           - (f_2^2 + 4\sin^4\psi)\cos^2\psi\right] ,
\eea
where $g(r)$ is now given by
\be
\label{gdef3}
 g(r) = r^2 h(r) \sqrt{\frac{\left(f_1^2+4\cos^4\psi\right)\left(f_2^2+4\sin^4\psi\right)}
 {\left(1+\frac{b^2}{r^4}\right) \left(1+ r^4 h(r) z'^2 + r^2 h(r)\psi'^2 -a'^2_0\right)} } \ .
\ee
There is also an equation of motion for $a_0(r)$, which can be integrated once to give
\be
\label{Aeom}
{g(r)\over h(r)}\left(1 + {b^2\over r^4}\right) a_0'(r) = \tilde{d}(r) \,, 
\ee
where $\tilde{d}(r) \equiv d - 2bc(r)$ is the radial electric displacement field at $r$,
and where $d$ is the integration constant. From the point of view of the boundary theory,
$d$ is the total charge density and $\tilde{d}(r)$ is the contribution to the charge density
coming from radial positions below $r$.
Clearly $\tilde{d}(\infty)=d$. In addition, $\tilde{d}(r_{min})$, if non-zero, corresponds to sources
at the bottom of the D7-brane.
The physical charge density is given by 
$D=(2\pi\alpha'/L)({\cal N}/V_{2,1})d$.

It is useful to decouple the $z(r)$ and $a_0(r)$ equations and express $g(r)$, and therefore $z'(r)$ and $a_0'(r)$,
just in terms of $\psi(r)$ and the constants $d$ and $c_z$:
\be
\label{geqn}
 g = {h\over \left(1+{b^2\over r^4}\right)}
  \sqrt{\frac{\tilde{d}^2+\frac{r^4}{h}\left(h\left(1+\frac{b^2}{r^4}\right)\left(f_1^2+4\cos^4\psi\right)\left(f_2^2+4\sin^4\psi\right)-\left(\frac{c_z}{r^4}-f_1 f_2\right)^2\right)}{1+hr^2\psi'^2}}  
\ee
\be
\label{zeqn}
 z' = \frac{\frac{c_z}{r^4}-f_1 f_2}{h}
 \sqrt{ \frac{1+hr^2\psi'^2}{\tilde{d}^2+\frac{r^4}{h}\left(h\left(1+\frac{b^2}{r^4}\right)\left(f_1^2+4\cos^4\psi\right)\left(f_2^2+4\sin^4\psi\right)-\left(\frac{c_z}{r^4}-f_1 f_2\right)^2\right) }   } 
\ee
\be\label{Aeqn}
 a'_0 = \tilde{d}\sqrt{ \frac{1+hr^2\psi'^2}{\tilde{d}^2+\frac{r^4}{h}\left(h\left(1+\frac{b^2}{r^4}\right)
 \left(f_1^2+4\cos^4\psi\right)\left(f_2^2+4\sin^4\psi\right)  -\left(\frac{c_z}{r^4}-f_1 f_2\right)^2\right)  } } \,.
\ee

\section{MN embeddings and quantum Hall states}\label{sec:qhstate}

MN embeddings are solutions with $\psi(\infty)=\psi_\infty$ and $\psi(r_0)=0$ or $\pi/2$ for
some $r_0>r_T$. In the boundary theory these embeddings describe states that have a mass-gap
for charged excitations.
This suggests that the MN embeddings describe electrical insulators.
As we will now argue, in the presence of a background magnetic field, they actually describe
quantum Hall states.

Smoothness of the embedding at $r=r_0$ requires $\psi'(r_0)\rightarrow \pm\infty$.
It also requires $z'(r_0)$ to be less singular than
$\psi'(r_0)$. This can be understood by looking at the behavior of the induced metric on the
D7-brane in the limit $r\rightarrow r_0$. 
For example, for the $\psi(r_0)=\pi/2$ case the relevant part of the induced metric is
\be
L^{-2}ds_{D7}^2 \approx \cdots 
+ \left(1 + {r^2 z^{\prime 2} + r^{-2} h(r)^{-1}\over \psi^{\prime 2}}\right) d\psi^2 
+(\frac{\pi}{2}- \psi)^2(d\Omega_2^{(1)})^2 \,,
\ee
which is non-singular as $r\rightarrow r_0$ as long as $z'$ is less singular than $\psi'$.
These conditions fix the integration constant in the MN embeddings to be $c_z=f_1 f_2 r_0^4$.

We also assume that the worldvolume gauge field is smooth at $r=r_0$, which requires that
\be
\label{locking}
\tilde{d}(r_0) = d - 2c(r_0) b = 0 \,.
\ee
This is equivalent to the requirement that there are no electric sources in the D7-brane.
Sources can be included by adding strings that stretch from the horizon to the D7-brane
and that are uniformly distributed in $(x,y)$.
However, these strings would tend to pull the D7-brane towards the horizon, leading to a BH embedding
(as in the supersymmetric D3-D7 system \cite{Kobayashi:2006sb}).
The entire charge density $d$ in the MN embedding is due purely to the CS term
and corresponds to a ``fluid" of D5-branes inside the D7-brane with a 
radial distribution given by $\tilde{d}'(r)$ and a uniform distribution in $(x,y)$.
We recognize (\ref{locking}) as the key property of a quantum Hall state, namely
that the fluid charge density and the magnetic field are locked.
The filling fraction is given by the ratio of the physical charge density to the physical magnetic field:
\be
\label{filling_fraction}
\nu = {2\pi D\over B} = {(2\pi\alpha')^2{\cal N}\over L V_{2,1}}{2\pi d\over b} = 
{N_3\over \pi} {d\over b} = {2N_3c(r_0)\over \pi} \,,
\ee
where, from (\ref{4-form_flux_2}) 
\be
c(r_0) =  \psi(r_0) - \psi_\infty + {1\over 4}\sin(4\psi_\infty)\,, \;\; \psi(r_0)=0 \; \mbox{or} \; {\pi\over 2}\,.
\ee

Smoothness of the gauge field also requires the absence of magnetic sources,
which correspond to a distribution of D5-branes ending on the D7-brane.
These too necessarily lead to a BH embedding.\footnote{We are grateful to Omid Hamid for pointing this out to us.}
The absence of magnetic sources implies that the magnetic flux on the $S^2$ that shrinks at $r=r_0$
must vanish.
For definiteness, we shall consider MN embeddings with $\psi(r_0)=\pi/2$, which means that $f_1=0$.
In this case
\be
f_{2}^2=4\sin^{2}\psi_{\infty}-8\sin^{4}\psi_{\infty} \,,
\ee
and the embeddings are stable for
\be
\cos^{2}\psi_{\infty}>\frac{48}{73} \,.
\ee
The condition $\Delta_{\pm}<0$ further restricts the range to 
\be
0.5235 \lesssim \psi_{\infty}  \lesssim 0.6251 \,.
\label{psirange}
\ee
The allowed values of $\psi_\infty$ are quantized, since the flux $f_2$ is quantized.

Thus, there is a discrete family of MN embeddings 
given by the allowed discrete values of $\psi_\infty$ in the range (\ref{psirange}).
We claim that each of these describes a specific quantum Hall state, with a 
filling fraction given by (\ref{filling_fraction}). 
The allowed filling fractions are discrete and lie in the range 
\be 
 0.6972\lesssim {\nu\over N_3} \lesssim 0.8045 \, . 
\ee
In addition, there are the embeddings with $\psi_\infty=0$ and $\psi_\infty=\pi/2$,
which, for $\psi(r_0)=\pi/2$, correspond to $\nu/N_3=1$ and $0$, respectively.

If we change the charge density relative to the magnetic field
and move away from the filling fraction (\ref{filling_fraction}),
the MN embedding deforms continuously into a BH embedding,
which, as we shall see in the next section, describes an ordinary conducting state.

%

\subsection{Conductivities}
\label{MNconductivity}

A quantum Hall state is a state of charged matter in 2+1 dimensions with a vanishing longitudinal
conductivity and a quantized transverse conductivity.
The existence of a mass gap in the MN embeddings, together with the quantized linear relation between the charge
density and magnetic field, suggest that they are in fact quantum Hall states.
We will verify this by computing directly the electrical conductivities.

To do this, we need to add more components to the D7-brane gauge field:
\be
A_x = {L\over 2\pi\alpha'}\left(te + a_x(r)\right) \; , \;
A_y = {L\over 2\pi\alpha'}\left(xb + a_y(r)\right) \; , \;
A_0 = {L\over 2\pi\alpha'} a_0(r) \,,
\ee
where $e$ is a (dimensionless) background electric field. The radial dependence of $a_x$ and $a_y$ will 
determine the longitudinal and transverse currents, respectively. The details of the action 
and equations of motion in this case can be followed in Appendix \ref{app:conductivity}.
The relevant results are expressions for the radial derivatives of the three gauge field components:
\bea
\label{A0appC}
 a'_0 & = & \left(h \left(1-\frac{e^2}{hr^4} \right) \tilde d +  \frac{eb}{r^4} \tilde j_y\right)\sqrt\frac{1+hr^4z'^2+hr^2\psi'^2}{X} \\
\label{A1appC}
 a'_x & = & j_x\left(1-\frac{e^2}{hr^4}
 +\frac{b^2}{r^4}\right)\sqrt\frac{1+hr^4z'^2+hr^2\psi'^2}{X} \\ 
 \label{A2appC}
 a'_y & = & \left(\frac{eb}{r^4} \tilde d-\left(1+\frac{b^2}{r^4}\right)\tilde j_y\right)\sqrt\frac{1+hr^4z'^2+hr^2\psi'^2}{X} \,,
\eea
where $j_x$ is the (dimensionless) longitudinal current (the constant of integration in the $a_x$ equation),
and $\tilde{j}_y$ is defined by analogy with $\tilde{d}$ as $\tilde{j}_y \equiv j_y - 2c(r)e$, 
where $j_y$ is the transverse current (the constant of integration in the $a_y$ equation).
As with the charge density, the physical currents are given by 
$J_i = (2\pi\alpha'/L)({\cal N}/V_{2,1})j_i$.
The quantity $X(r)$ is given by
\bea
\label{X(r)}
X(r)   & = & h\left(1+\frac{b^2}{r^4}-\frac{e^2}{hr^4}\right)\left(hr^4(f_1^2+4\cos^4\psi)(f_2^2+4\sin^4\psi)+h\tilde d^2 - \tilde j_{y}^{2} - j_x^2\right) \nonumber \\ 
   &   & -(\sqrt h b \tilde d-e \tilde j_y)^2 \ .
\eea
Regularity of the three gauge field field components at $r=r_0$ requires both $\tilde{d}(r_0) = 0$ and
$\tilde{j}_y(r_0)=0$, as well as $j_x=0$.
The physical conductivities are therefore given by
\be
\sigma_{xx} = {J_x\over E} = 0 \,
\ee
and
\be
\sigma_{xy} = {J_y\over E} = {N_3\over 2\pi^2} {j_y\over e}
= {N_3 c(r_0)\over \pi^2} = {\nu\over 2\pi} \,.
\ee
The MN embeddings, when they exist, describe quantum Hall states
with quantized transverse conductivities.
Furthermore, in the holographic description, the quantization is topological since it
originates from the Dirac quantization of the magnetic fluxes on the $S^2$'s.
In particular, $\sigma_{xy}$ in the MN embeddings is independent of the temperature.

\subsection{Numerical results}

\begin{figure}[ht]
\center
\includegraphics[width=0.40\textwidth]{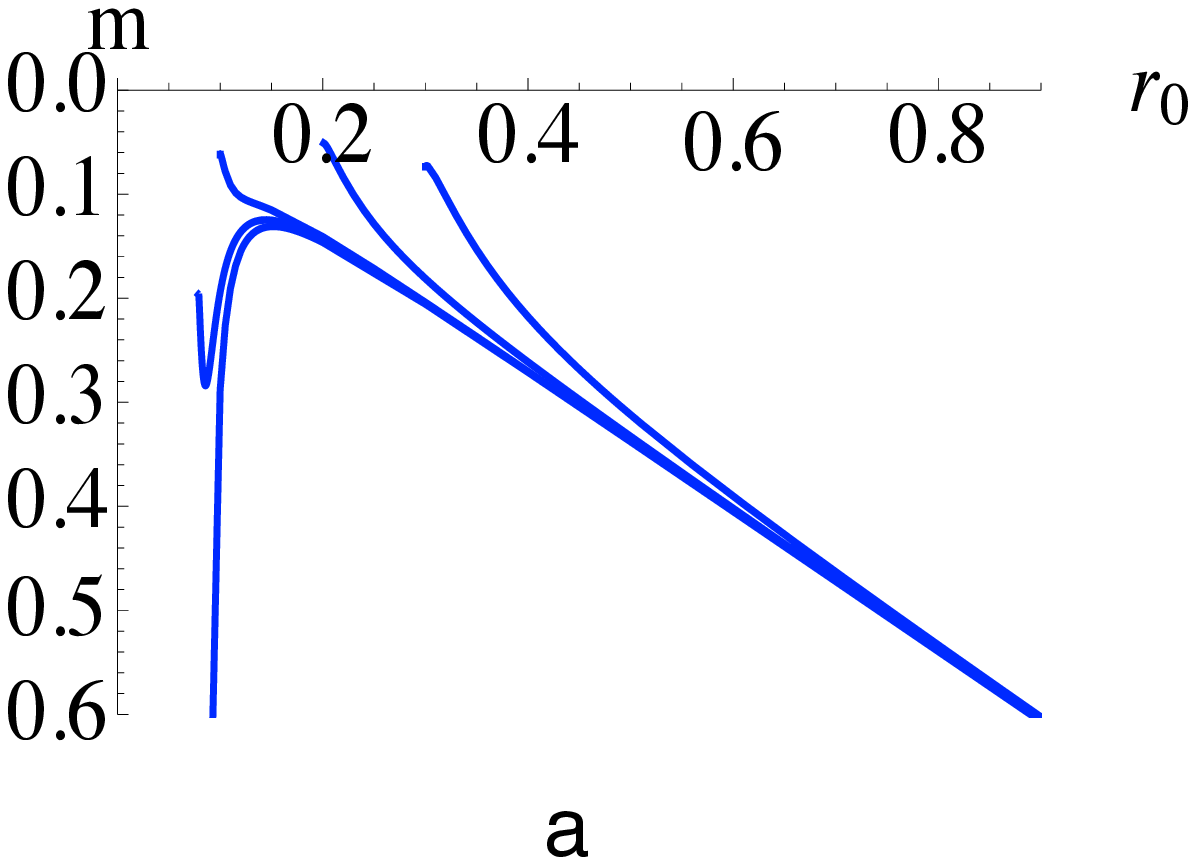}
\hspace{1cm}
\includegraphics[width=0.40\textwidth]{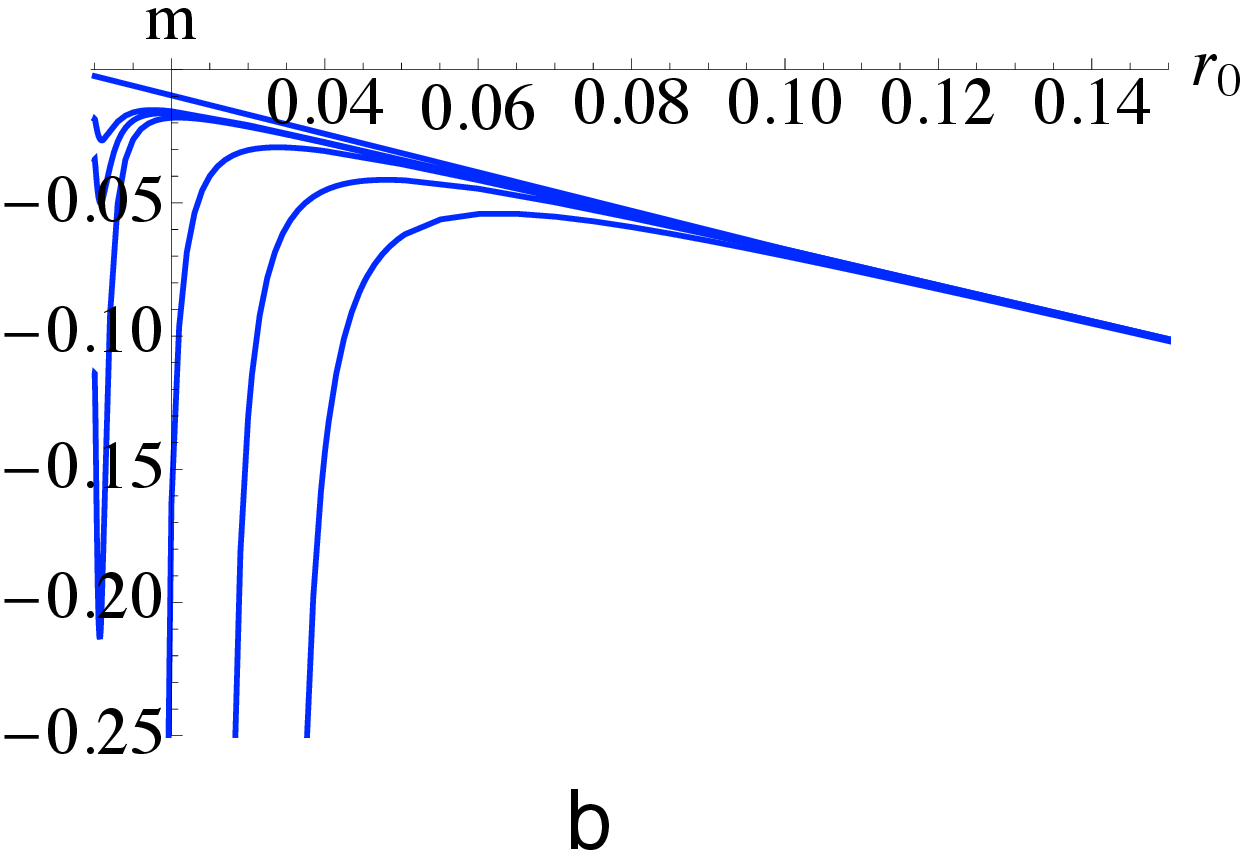} 
\caption{MN embeddings: (a) at fixed $d=0.01$, $\Delta_{+}=-1$, and with
$r_{T}=0, 0.079, 0.1, 0.2, 0.3$ from bottom to top, and (b) at fixed 
$r_{T}=0.05734$, $\Delta_{+}=-1$, and with
$d=0,0.00015,0.00017,0.0002,0.0005,0.001,0.0017$ from top to bottom.}
\label{fig:massr0}
\end{figure}

We find numerical solutions by shooting from some $r_{0}$ with initial conditions
$\psi(r_0)=\pi/2$ and $\psi'(r_0)\rightarrow -\infty$ and then extracting the 
parameter $m$ from the large $r$ behavior.
Figure \ref{fig:massr0} shows $m$ as a function of $r_0$ for different temperatures and different
values of the magnetic field $b>0$ (and therefore of the charge density $d$).
Note that all the solutions have $m<0$. 
Since the equations of motion are invariant under the transformation
\be
\label{mirror_symmetry}
f_1\leftrightarrow f_2 \; , \; b\rightarrow -b \; , \; \psi \rightarrow {\pi\over 2} - \psi \,,
\ee 
there are ``mirror" solutions with $m>0$ and $b<0$ (and with $f_2=0$ and therefore $\psi(r_0)=0$).
MN embeddings with equal sign $m$ and $b$ appear to be excluded.
It would be interesting to understand this from the point of view of the boundary theory.\footnote{In three dimensions
the sign of the fermion mass is related to its spin. A possible interpretation of the above observation is that 
the gapped state only exists when the spins are aligned with the magnetic field.}

\begin{figure}[ht]
\center
\includegraphics[width=0.40\textwidth]{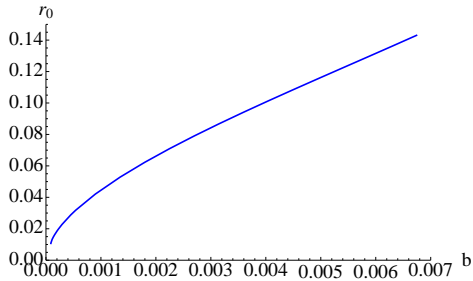}
\caption{The mass gap ($r_0$) as a function of the magnetic field (at $r_T=0.01$).
Fitting to $r_0 = \alpha + \beta \sqrt{b} + \gamma \, b$ we find 
$\alpha = 5.695\times 10^{-6}$, $\beta = 1.176$, and
$\gamma = 6.719$.}
\label{r0vsb}
\end{figure}

At low temperatures such that $r_{T} \ll \sqrt{b}$ there are two solutions above a critical value of $|m|$,
which increases as $b$ (and with it $d$) increases.
As the temperature is increased (at fixed $b$ and $d$), or alternatively as the magnetic field $b$ 
(and accordingly $d$) is decreased (at fixed temperature), the solution 
at smaller $r_0$ ceases to exist above another
critical $|m|$, and a third solution with an even smaller $r_0$ appears below this value of $|m|$.
These two solutions 
disappear completely above some maximal temperature, whereas 
the solution with the largest $r_0$ appears to exist at all temperatures.
In Section \ref{sec:thermo} we will study the thermodynamics of the system;
the small $r_0$ embeddings are thermodynamically preferred to the large $r_0$ embedding
and there is a first order phase transition to a BH embedding at a critical temperature, which is below the 
above maximal temperature.
We can also see in Fig.~\ref{fig:massr0} that $r_0$, and therefore the size of the mass gap,
increases with $b$ in the relevant MN embedding (see Fig.~\ref{r0vsb}).

\section{BH embeddings}\label{sec:metallic}

BH embeddings are solutions in which the D7-brane crosses the horizon at $r=r_T$
and therefore correspond to gapless ``metallic" states.
Since $h(r_T)=0$, consistency of the $z(r)$ equation (\ref{zeom2}) fixes the
integration constant $c_z = f_1 f_2 r_T^4$. 
These embeddings generically exist for any $f_1, f_2$ satisfying the condition that 
$\Delta_{\pm}$ are real and for any values of $d, b$ and $m$.
Relative to the MN embeddings, there are two additional parameters $\tilde{d}(r_T)=d-2c(r_T)b$ 
and $f_1$, corresponding to the possibility of adding electric and magnetic sources.

\subsection{Conductivities}\label{sec:BHconductivity}

Using the now standard Karch-O'Bannon technique \cite{Karch:2007pd, O'Bannon:2007in}, 
we can compute the currents by requiring reality of the action or the equations of motion
(see also \cite{perkraus}).
As before, the details can be found in Appendix C.
In particular, the quantity $X(r)$ in (\ref{X(r)}) must be non-negative.
We define the ``pseudo-horizon" radius $r_*$ as the value of $r$  where the second factor in the
first term in (\ref{X(r)}) vanishes, {\em i.e.},
\be
\label{r*eqn}
e^2   =  h(r_*)(r_*^4 + b^2) \,.
\ee
Non-negativity of $X(r)$ then requires the second term, as well as the 
third factor in
the first term, to vanish at $r=r_*$. This gives the following conditions on the currents:
\bea
\label{Halleqn}
e \tilde{j}_y(r_*) &=& b\, h(r_*) \tilde{d}(r_*) \\
 \label{longitudinaleqn}
 \left(\tilde{j}_y(r_*)\right)^2+j_x^2 &=& h(r_*)\left(r_*^4 \left(f_1^2+4\cos^4\psi(r_*)\right)
 \left(f_2^2+4\sin^4\psi(r_*)\right)+\tilde{d}(r_*)^2 \right) .
\eea
The conductivities can be extracted in the linear response approximation, where 
we expand to the lowest non-trivial order in the electric field $e$.
In this approximation, equation (\ref{r*eqn}) gives
\be 
 r_*^4 \approx r_T^4\left(1+\frac{e^2}{b^2+ r_T^4}\right)  \;\; , \;\;
 h(r_*)\approx \frac{e^2}{b^2+r_T^4} \,.
\ee
The transverse conductivity is determined from (\ref{Halleqn}):
\be
\label{sigmaxy_BH}
\sigma_{xy} = {N_3\over 2\pi^2} {j_y\over e} = 
{N_3\over 2\pi^2}\left(\frac{b}{b^2+r_T^4}\tilde{d}(r_T)+2c(r_T)\right) \,,
\ee
and longitudinal conductivity is obtained from (\ref{longitudinaleqn}):
\be
\label{sigmaxx_BH}
\sigma_{xx} = {N_3\over 2\pi^2} {r_T^2\over b^2 + r_T^4}
\sqrt{\tilde{d}(r_T)^2 + (f_1^2+4\cos^4\psi(r_T))(f_2^2+4\sin^4\psi(r_T))(b^2+r_T^4)}\,.
\ee
As expected for a metallic state, both the longitudinal and transverse conductivities are non-vanishing.
The transverse conductivity has two components: the first comes from a dissipative system of charges
at the horizon with charge density $\tilde{d}(r_T)$, and the second from a dissipationless charge fluid with charge density
$d-\tilde{d}(r_T) = 2c(r_T)b$. The longitudinal conductivity involves only the first component,
and the rest corresponds to pair production. (See \cite{Lifschytz:2009si} for a discussion of the same behavior in the 
Sakai-Sugimoto model.)
At $T=0$, $\sigma_{xx}=0$ and $\sigma_{xy} = \frac{2\pi D}{B}$, as required by Lorentz invariance. 
At high temperature, the longitudinal conductivity approaches a constant value.
The high-temperature behavior of the transverse conductivity depends on the specific solution
and can be determined numerically.

\subsection{Numerical results} 

For a given $f_1,f_2,d,b$ and temperature, the solution is parameterized by the value of $\psi$ at the
horizon, $\psi(r_T)$. The value of the derivative at the horizon $\psi'(r_T)$ is fixed by $\psi(r_T)$
and should not be imposed as a separate boundary condition (see Appendix~\ref{app:nearhorizon}).
The equation of motion for $\psi$ (\ref{psieom3}) 
is solved by shooting from different values at the horizon $\psi(r_T)$,
and then $m$ is determined as a function of $\psi(r_T)$.
As for the MN embeddings, the symmetry (\ref{mirror_symmetry}) implies that there are mirror
embeddings with opposite $b$ and $m$, and with $f_1$ and $f_2$ exchanged.

\begin{figure}[ht]
\center
\includegraphics[width=0.50\textwidth]{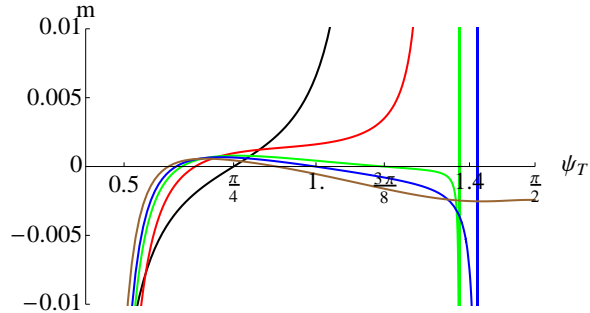}
\caption{
BH embeddings at fixed $r_{T}=0.01$, $d=0$, $b=0$, $\Delta_+=-1$,
and with $f_1=1/\sqrt{2}$ (black), $0.5387$ (red), $0.4$ (green), $0.308$ (blue), and $0$ (brown).}
\label{fig:masspsiT}
\end{figure}

There are several parameters which we can vary.
We are primarily interested in the parameter subspace given by $f_1=0$, which is
where the MN (quantum Hall state) embeddings appear. 
However, it is instructive to consider more general BH embeddings with $f_1\neq 0$.
%
Figure \ref{fig:masspsiT} shows $m$ as a function of $\psi(r_T)$ for a number of solutions with $d=b=0$, 
$r_T=0.01$, and different values of $f_1$ and $f_2$, but with fixed exponents $\Delta_{\pm}$.
It is clear that BH embeddings exist at all values of $m$.
We start with $f_1=f_2$ (black curve, left-most for positive $m$) and gradually decrease $f_1$ and increase $f_2$.
Initially there is a single solution for each value of $m$, but at some point more solutions appear for some values of $m$.
As $f_1\rightarrow 0$ (brown curve, flat near $\psi(r_T)=\pi/2$) some of the solutions are harder to see, since
they get squeezed into $\psi(r_T)=\pi/2$. In particular, there is a solution for arbitrarily large positive $m$.

\begin{figure}[ht]
\center
\includegraphics[width=0.350\textwidth]{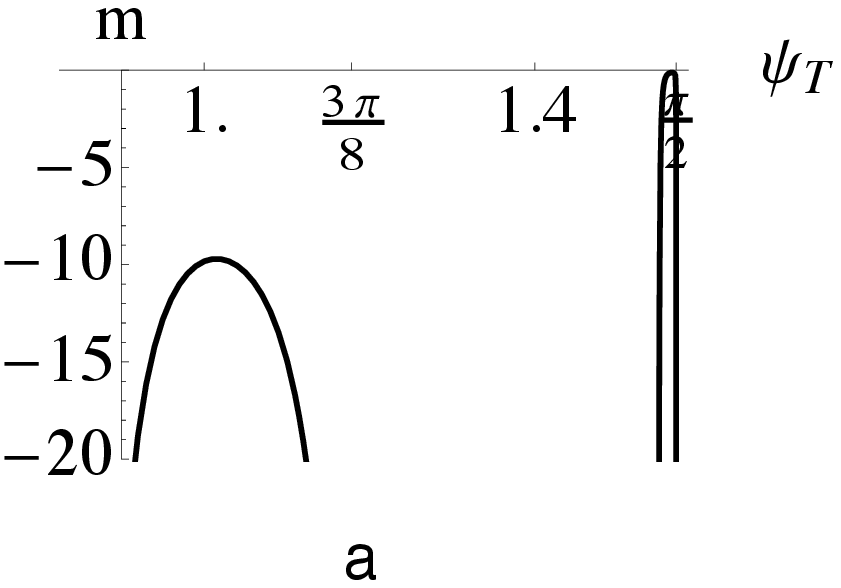} 
\hspace{1cm}
\includegraphics[width=0.350\textwidth]{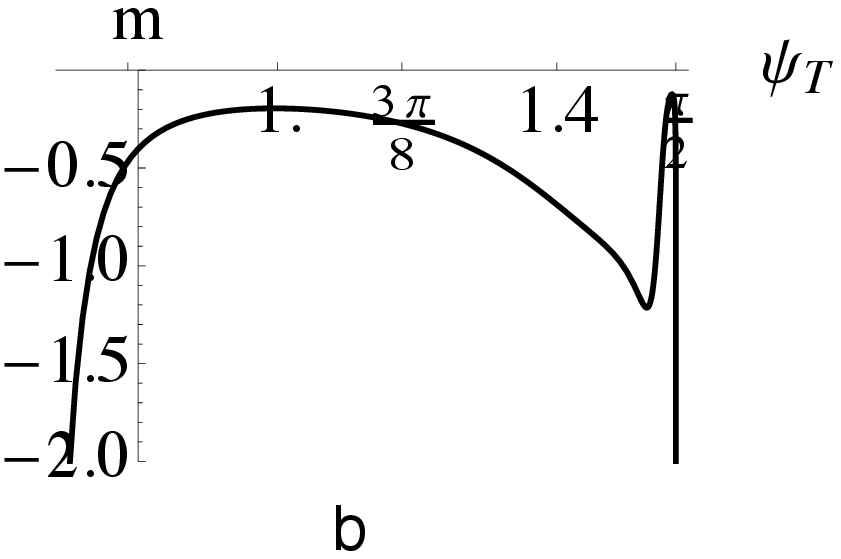} 
\caption{BH embeddings with $f_1=0$, 
$d=0.01$, $b=0.0044$, $\Delta_{+}=-1$, and (a) $r_{T}=0.06$ (b) $r_T=0.07$.}
\label{fig:masspsiT1}
\end{figure}

\begin{figure}[ht]
\center
\includegraphics[width=0.30\textwidth]{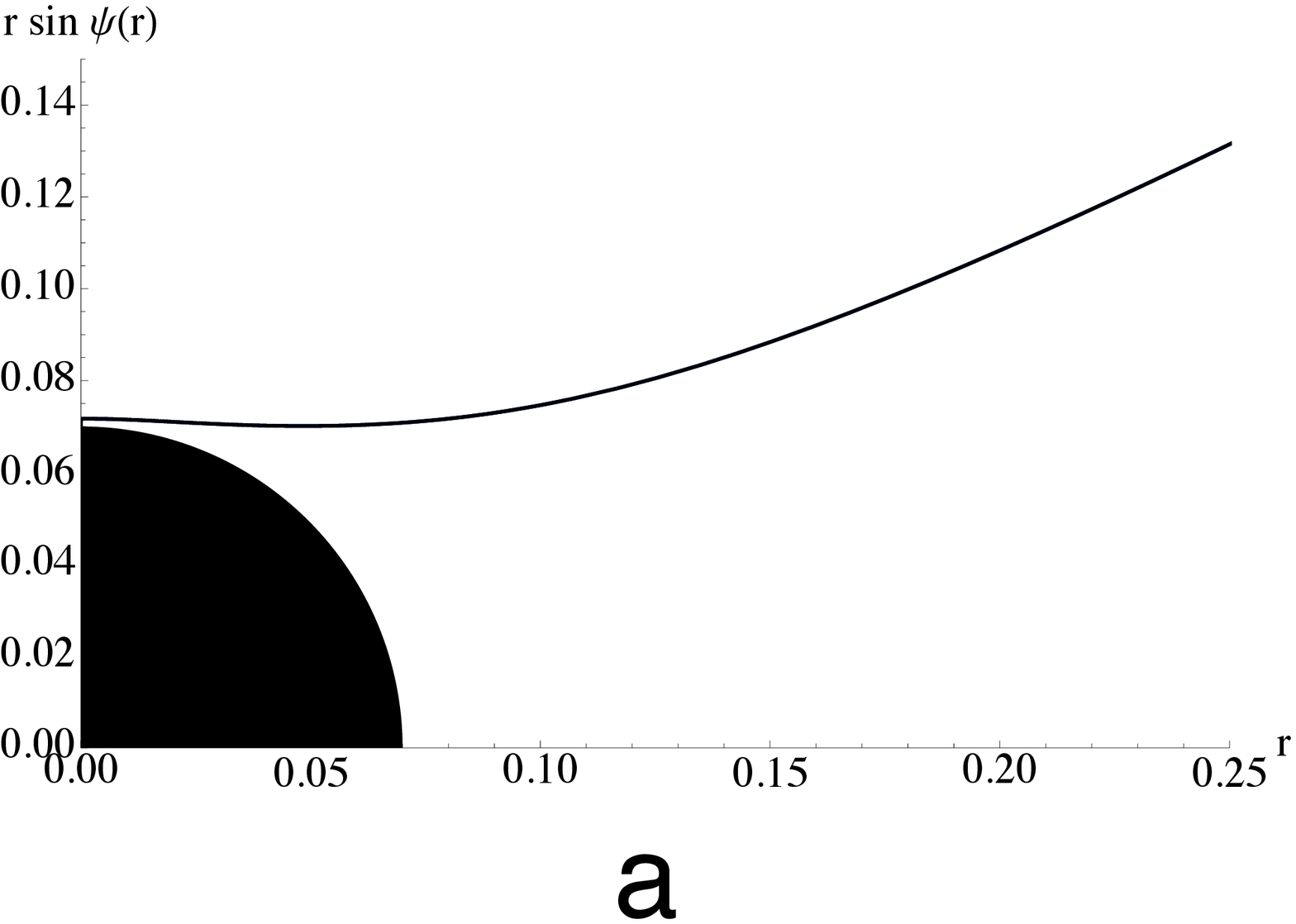}
\hspace{0.5cm}
\includegraphics[width=0.30\textwidth]{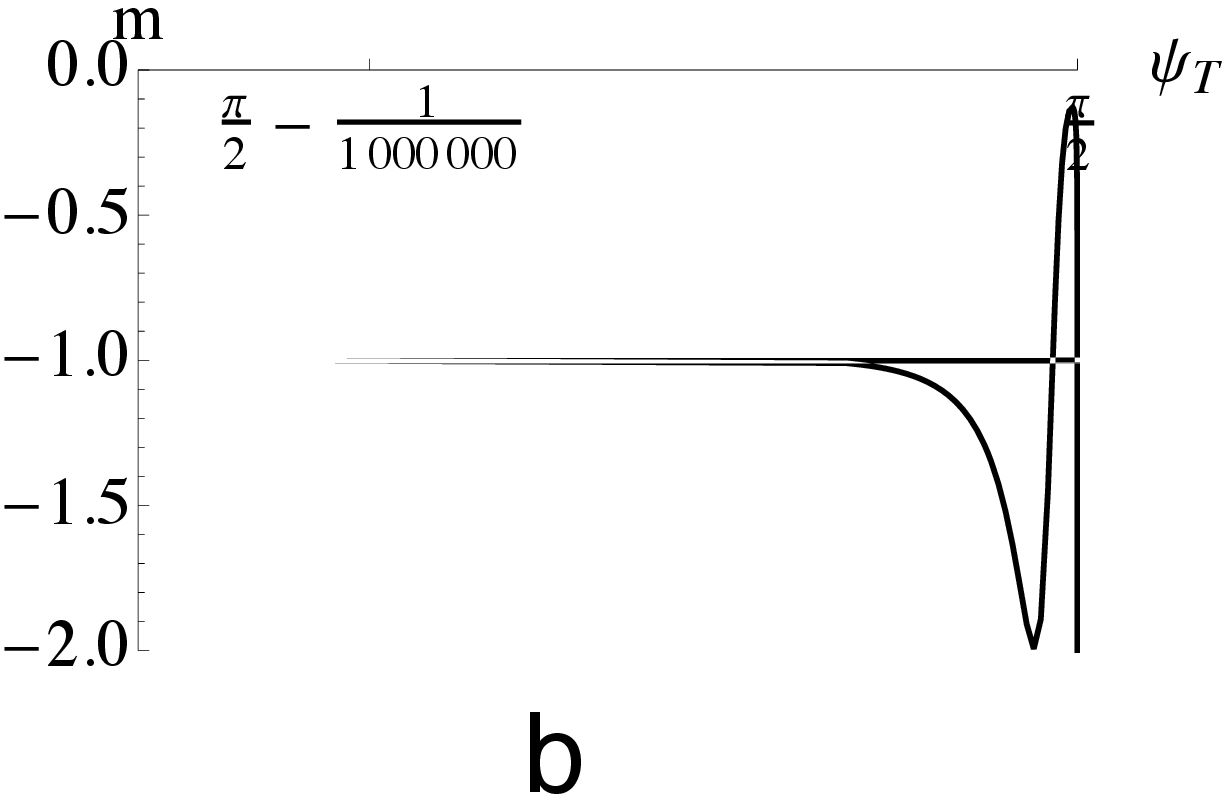} 
\hspace{0.5cm}
\includegraphics[width=0.30\textwidth]{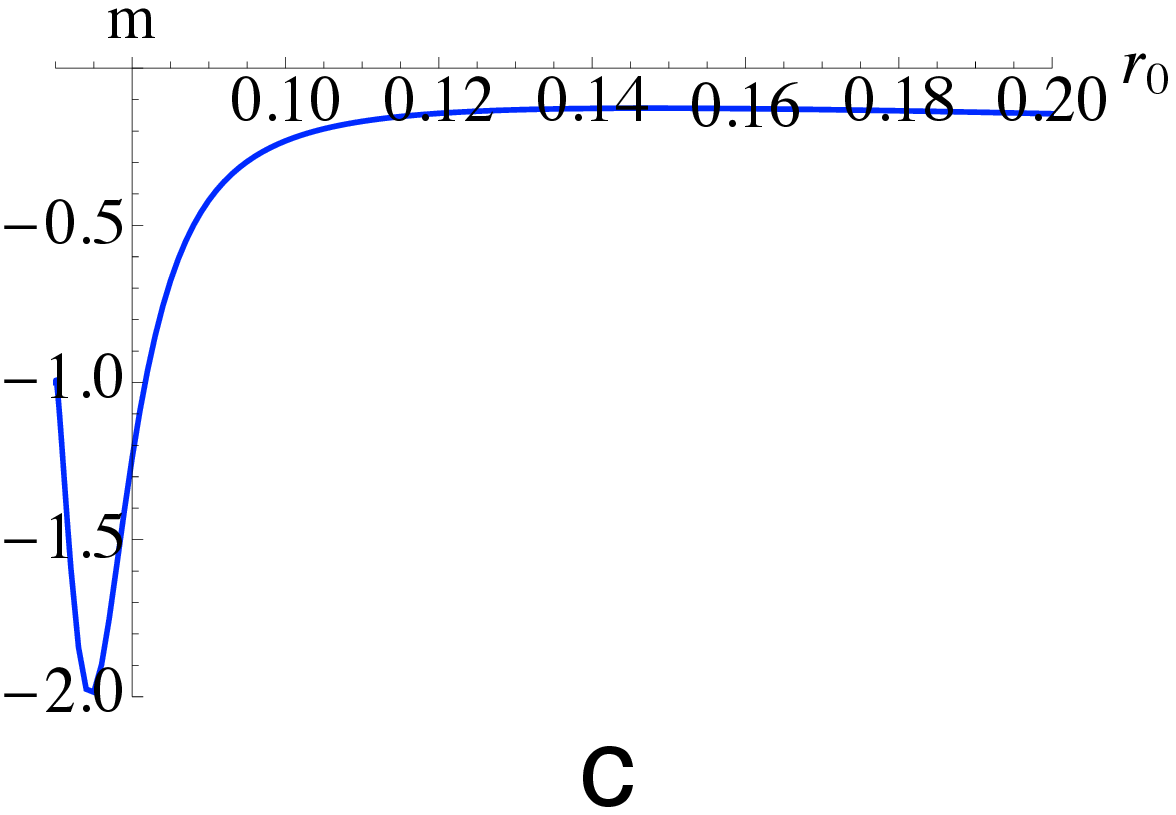} 
\caption{Comparing the ``spiky" BH embeddings for small $\tilde{d}(r_T)$ with the MN embeddings:
(a) MN and BH embeddings superimposed. (b) $m$ vs. $\psi(r_T)$ for the BH embeddings.
(c) $m$ vs. $r_0$ for the MN embeddings.}
\label{BH_MN_match}
\end{figure}

Let us now focus on the case $f_1=0$ and consider non-zero values of $d$ and $b$, but close to
the source-free case, {\em i.e.}, with a small $\tilde{d}(r_T)=d-2c(r_T)b$.
The result shown in Fig.~\ref{fig:masspsiT1} reveals as many as four solutions, depending on the range of $m$.
Note that, like the MN embeddings, these BH embeddings exist only for $m<0$.
We also observe that two of the solutions come arbitrarily close to $\psi(r_T)=\pi/2$ as
we reduce $\tilde{d}(r_T)$.
These solutions correspond to ``spiky" embeddings that enter the horizon very close to $\psi(r_T)=\pi/2$
but otherwise closely resemble the MN embeddings (see Fig.~\ref{BH_MN_match}a).
Note also that the longitudinal conductivity in these cases (\ref{sigmaxx_BH}) goes to zero as 
$\tilde{d}(r_T)\rightarrow 0$ and $\psi(r_T)$ approaches $\pi/2$, which agrees  
with this property of the MN embeddings.
We can therefore identify these two BH embeddings as the result of adding electric sources 
to the two main MN embeddings.
In fact we find a perfect match between the BH embeddings close to $\psi(r_T)=\pi/2$
in the limit $\tilde{d}(r_T)\rightarrow 0$
and the MN embeddings, including the third solution at the smallest value of $r_0$
(Fig.~\ref{BH_MN_match}b,c).
%

However, this cannot be the complete set of solutions.
Since we had only one solution when $f_1=f_2$, there must be an odd number of solutions at any given $m$.
Furthermore, we saw that there is a solution for arbitrarily large positive $m$.
There must therefore be at least one more solution that extends to $m\rightarrow\infty$ very close to 
$\psi(r_T)=\pi/2$. This solution is very hard to see numerically when $f_1=0$.

\section{Thermodynamics}\label{sec:thermo}

To determine which state is preferred thermodynamically, 
we should compare the free energies of the different solutions.
We will work in an ensemble where the parameters $(f_{1}, f_{2}, m, b,d)$ are fixed. There may be other ensembles which are relevant but we will not consider them here.
The Euclidean D7-brane action, evaluated for a BH or an MN embedding solution, defines the grand canonical
potential (density) for the corresponding state,
\be 
\Omega(\mu,T,b) = {1\over {\cal N}} S_{D7}^E[\psi(r),z(r),a_0(r)]_{solution} \,.
\ee
The chemical potential is defined, as usual, by $\mu = a_0(\infty)$.
For MN embeddings, the RHS includes the boundary term
$ - 2 c(r_0) b a_0(r_0) = - d a_0(r_0)$ from the Euclidean version of the boundary action (\ref{boundary_action}).
For BH embeddings the boundary action vanishes since $a_0(r_T)=0$.
The bulk part of the action depends only on $a_0'(r)$ and is therefore naturally
expressed in terms of the charge density $d$.
To define the grand canonical potential we therefore need to solve for $d(\mu)$.
In the case of BH embeddings, the chemical potential is determined uniquely by the
charge density, since
\be
\label{mu_BH}
\mu = \int_{r_{T}}^\infty dr a_0'(r) \,.
\ee
The RHS is a function of $d$, which we can numerically invert to obtain $d(\mu)$. 
For MN embeddings, the charge density is fixed to $d=2c(r_0)b$
and is therefore independent of $\mu$.
In this case
\be
\mu - a_0(r_0) = \int_{r_{0}}^\infty a_0'(r) \,,
\ee
so the difference $\mu - a_0(r_0)$ is fixed by $b$ and $c(r_0)$.
The necessity of the boundary term can now also be understood from the condition that
$d= - \partial\Omega/\partial\mu$.

The bulk part of the action is divergent at large $r$ for both the MN and BH embeddings
and should be regulated using holographic renormalization.
The holographic counterterms one needs depend on $\Delta_{+}$.
For $-3/2<\Delta_{+}<-1$ the only counterterm is
\be
 S_{1,\rm{counter}} = -\frac{{\cal N}}{6} \int d^3x\, \sqrt{\gamma}\sqrt{1+\frac{1}{2}(2\pi \alpha^{'}F_{\mu\nu})^2}\  
 \left(2+\frac{3+2\Delta_{+}^{2}\Delta_{-}}{3+2\Delta_{+}}(\psi-\psi_{\infty})^2\right)\,,
\ee
where $\gamma$ is the determinant of the induced metric on the surface at some UV cutoff (usually taken to be some $r_{max}$).
For $\Delta_+\geq -1$ there are additional counterterms.
In particular, for $\Delta_{+}=-1$ there is an additional logarithmic counterterm
\be
S_{2,\rm{counter}} = -{\cal N}\int d^3x\sqrt{\gamma}(\psi-\psi_{\infty})^3\frac{2\sin 4\psi_{\infty}(f_{1}^2+f_{2}^{2}+4-10\sin^{2} 2\psi_{\infty})\ln \Lambda}{(f_{1}^{2}+4\cos^4 \psi_{\infty})(f_{2}^{2}+4\sin^4 \psi_{\infty})-f_{1}^{2}f_{2}^{2}} \,,
\ee
where $\Lambda$ is the cut-off.
%

\begin{figure}[ht]
\center
\includegraphics[width=0.7\textwidth]{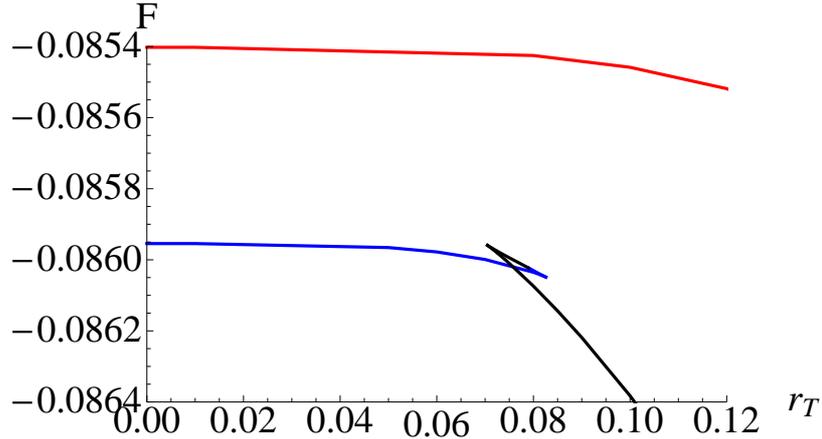} 
\caption{Free energy as a function of the temperature at $f_1=0$, $d=0.01$,  $\Delta_{+}=-1$ and $m=-0.2$:
BH embeddings in black, small $r_0$ MN embeddings in blue, and large $r_0$ MN embedding in red.}
\label{fig:thermophase}
\end{figure}

Since the action is already expressed in terms of the charge density, it is somewhat simpler
to work in the canonical ensemble. The free energy is defined by
\be
F(d,T,b) = \Omega(\mu,T,b) + \mu d \,.
\ee
We compare the free energies of the MN embeddings and the BH embeddings with $f_1=0$
and $\tilde{d}(r_T)=0$. Figure \ref{fig:thermophase} shows our results.
The relevant embeddings are the two MN embeddings with the smaller $r_0$ (flatter blue curve),
and the two BH embeddings with the smaller $\psi(r_T)$ (steeper black curve).
The figure shows the typical behavior of a system that undergoes a first order phase transition.
Below a critical temperature, the MN embedding with the larger $r_0$ of the two is preferred, 
and the system is in a quantum Hall state.
Above this temperature the BH embedding with the smaller $\psi(r_T)$ of the two is preferred, 
and the system is in a metallic state. The other relevant MN embedding and BH embedding
are the unstable states that make up the ``swallow tail" in the figure.
These embeddings meet at a critical embedding that touches the horizon.
We have also included the free energy of the MN embedding with the largest $r_0$ (upper red curve).

\section{Discussion}

In this paper we have exhibited quantum Hall states in a system of strongly coupled
fermions in 2+1 dimensions.
Specifically, we used gauge/gravity duality to study a system of electrically charged fermions 
in $2+1$ dimensions which interact strongly with a $3+1$ dimensional large $N$ non-abelian 
gauge field, at finite temperature, charge density, and in the presence of a background magnetic field.  
The holographic dual description consists of a probe D7-brane in the near-horizon background
of non-extremal D3-branes, with certain worldvolume fluxes turned on for stability.
We found embeddings that avoid the horizon, which we identified as quantum Hall states,
with a variety of discrete filling fractions, and embeddings that enter the horizon,
which we identified as metallic states.
We computed the longitudinal and transverse conductivities in each case and confirmed
these identifications. 
As the magnetic field is varied for fixed charge, the quantum Hall states turn into
metallic states in a continuous 
fashion. However, as a function of the temperature there is a first-order phase transition from a quantum 
Hall state at low temperature to a metallic state at high temperature.

While this framework can be used to study certain properties of the strongly-coupled quantum Hall
state that we found, such as electric transport, excitations, and hydrodynamics,
there is still much to be understood before we can conclude that a true fractional
quantum Hall effect has been exhibited. Let us list some important open issues
and suggest some ideas about how one might try to resolve them.

\medskip

\noindent 1. \underline{\em Charge quantization and quasiparticles}:

\smallskip

\noindent The first issue has to do with the basic unit of charge.
From the point of view of the D7-brane worldvolume theory, there are $N_3$
``flavors" of charged fermions, each carrying a unit of charge under the D7-brane
worldvolume gauge field.
However, it is not immediately clear what should be regarded as the ``electron" for this system.
We would like to suggest that the ``electron"
be identified with the $SU(N_3)$ singlet baryon state,
described by a D5-brane wrapped on $S^5$, with $N_3$ strings connecting it to the 
D7-brane. This would mean that the basic unit of charge is $N_3$ times that of
the fundamental fermions.
It seems natural to associate the quasiparticles with fractional baryons, 
which are described by D5-branes that wrap a part of the $S^5$ and end at the horizon
(see for example \cite{Callan:1998iq}). 
These D5-branes have a smaller number of attached strings and would therefore correspond to 
fractionally charged objects from the point of view of the D7-brane worldvolume.
However, at this point it seems that any fraction would be possible, depending on how the D5-brane 
is embedded, and in particular where it enters the horizon.


\medskip

\noindent 2. \underline{\em Different filling fractions and transitions}:

\smallskip

\noindent We are able to realize different filling fractions by turning on different
values of $f_2$. Since the fluxes are quantized and bounded, 
the set of allowed filling fractions is discrete and also bounded.
Generically, the filling fractions we get are not simple rational numbers as in
the real FQHE, but we can get close to any rational number by taking 
the flux quanta $n_2$, and the curvature radius in string units $L/\sqrt{2\pi\alpha'}$, 
to be arbitrarily large.

The main question is how to describe transitions between quantum Hall states with 
different filling fractions.
With several quantum Hall states at a given value of $L$, it is still not
clear how to describe transitions between them, since
the corresponding D7-brane embeddings differ at large $r$.
The fluxes are different and uniform in $r$, and correspondingly, the asymptotic
angles $\psi_\infty$ are different.
Since this is a non-normalizable mode, it corresponds to an external parameter in the 
boundary theory and does not change dynamically.
However, as we know from the case of the chemical potential and charge density,
there should be an alternative ensemble where the parameter (like $\mu$) becomes dynamical
and its dual dynamical variable (like $d$) becomes a parameter.
If we could identify the dual variable to $\psi_\infty$, this would then allow us
to define a new ensemble, in which $\psi_\infty$, and therefore the filling fraction, was
a dynamical variable, to be determined by minimizing the (appropriately Legendre transformed)
action.

\medskip

\noindent 3. \underline{\em Impurities and plateaus}:

\smallskip

\noindent 
The real quantum Hall effect exhibits finite plateaus around the quantized filling fractions,
so that there is a range of magnetic fields, at fixed density, in which the system remains
in the same quantum Hall state. The plateaus are due to the presence of impurities,
which give rise to localized states, that do not contribute to the conductivities.
In order to ``broaden" our quantum Hall states into plateaus, we need to be able to vary
$b$ without changing $d$, or vice versa, in the MN embedding.

Let us suggest a possible way to do this.
We can imagine changing the charge density without changing the magnetic field
by adding strings 
distributed in the $(x,y)$ plane that end on the D7-brane.
However, these strings cannot come from the horizon, since, at any non-zero density,
they would lead to a BH embedding.
What we need is a configuration of strings, such that for small non-zero densities,
the D7-brane remains in an MN embedding outside the horizon.
This might be achieved by the introduction of additional branes intersecting
the D7-brane, and localized in the $(x,y)$ plane. 
These ``impurity" branes add localized charged states to the system,
which correspond to strings between them and the D7-brane. By exciting these states
on a finite density of impurity branes, we could vary $d$ without changing $b$.
On the other hand, these impurity branes contribute to the free energy of the system,
and at a high enough density the BH embedding may be preferred.
The net effect of the impurity branes would thus be to ``postpone" the 
deformation of the MN embedding into a BH embedding
as the charge is varied, and thereby
broaden the quantum Hall state into a plateau.
Combining this proposal with the previous one, for including the different quantum
Hall states, would produce a series of transitions,
as the magnetic field, or charge density, is varied, from the metallic state to
a quantum Hall plateau, to the metallic state, to another quantum Hall plateau, and so on.

\medskip

\noindent 4. \underline{\em Edge states}:

\smallskip

\noindent Quantum Hall states in bounded regions exhibit
gapless chiral-fermionic edge excitations.
In the integer case these 1+1 dimensional edge states are related to the topological nature of the
quantum Hall state.
For example, in the annulus geometry the quantized filling fraction is given by 
the number of chiral excitations on one boundary minus the number of anti-chiral
excitations on the other.
To model an edge in our holographic model requires the D7-brane to somehow end
in one of the spatial directions $x$ or $y$.
Strictly speaking, this is only possible if we have a D9-brane, which we cannot add
because of the RR tadpole condition. 

Putting aside the issue of the edge for a moment, we can model 1+1 dimensional
chiral fermions by a second D7-brane, which now wraps the entire $S^5$,
and extends along $t,r$, and only one of the spatial directions, say $x$.
The new D3-D7 strings have $\# ND=8$, and describe massless chiral fermions
on the line $x$.
The worldvolume CS term on the new D7-brane, now more conveniently expressed as
$\int F_5\wedge  A\wedge F$,
will then include a term of the form $\int dr \, a_x(r)a_0'(r)$,
implying that a nonzero charge 
generates a current in the $x$ direction.

This might lead to a relation between the quantization of the filling fraction
and chiral edge states in our model.
The configuration with the two different D7-branes has an instability due to
a tachyon at their intersection.
The condensation of this tachyon will lead to a single D7-brane, which is embedded
in a more complicated way, and could potentially describe the system with the
chiral edge states.

\bigskip
\noindent

{\bf \large Acknowledgments}

We thank Matti J\"arvinen, Esko Keski-Vakkuri,
Sean Nowling, Daniel Podolsky, Danny Shahar, Efrat Shimshoni, and especially 
Ady Stern for useful comments and discussions. We also thank Omid Hamid for pointing out an error 
in an earlier version.
N.J. has been supported in part by the Israel Science Foundation under grant no.~392/09 and in part at the Technion
by a fellowship from the Lady Davis Foundation. N.J. wishes to thank University of Crete and CERN for hospitality while this 
work was in progress.  The research of M.L. is supported in part by the Israel Science Foundation under grant 
no. 568/05 and in part by the European Union grant FP7-REGPOT-2008-1-CreteHEPCosmo-228644.  
M.L. would like to thank the Univeristy of California, Berkeley and CERN for their hospitality.
The work of O.B. and G.L. was supported in part by the Israel Science Foundation under grant no.~392/09.
O.B. also thanks the Princeton Center for Theoretical Science for hospitality.

\appendix

\section{Large $r$ asymptotic analysis}
\label{app:stability}

In this appendix, we analyze the large $r$ behavior of $\psi$ in order to find the conditions on $f_1$, $f_2$, and $\psi_\infty$ required both for stability and for a dual interpretation as a fermion mass.  
We assume that $0 < \psi_\infty < \pi/2$.
Although we do not include here the effects of a nonzero charge density or background magnetic field, these contributions are subleading at large $r$, and so play no role in this discussion.

As before, consider the ansatz $\psi\to \psi_\infty + Ar^{\Delta}$, where ${\rm Re} \ \Delta<0$.  To leading order in $r$, the $\psi$ equation of motion (\ref{psieom3}) gives
\be
\label{psieomzerothorder}
f_1^2 \sin^2\psi_\infty-f_2^2 \cos^2\psi_\infty+4\sin^2\psi_\infty\cos^2\psi_\infty(\cos^2\psi_\infty-\sin^2\psi_\infty)  = 0 \ ,
\ee
which, for any $f_1$ and $f_2$, implicitly gives the value of $\psi_\infty$. For the case $f_1=f_2$, this implies $\psi_\infty = \pi/4$.  Expanding equation (\ref{psieom3}) to next order gives
\be
\Delta(\Delta+3) = 4 \sin\psi_\infty\cos\psi_\infty  \frac{f_1^2 + f_2^2 + 4 \left(\cos^4\psi_\infty - 4\sin^2\psi_\infty\cos^2\psi_\infty + \sin^4\psi_\infty\right)}{f_1^2\sin^4\psi_\infty +  f_2^2\cos^4\psi_\infty + 4\sin^4\psi_\infty\cos^4\psi_\infty} \ .
\ee
Using (\ref{psieomzerothorder}) to solve for $f_2$, we can solve for $\Delta$ in terms of $f_1$ and $\psi_\infty$:
\be
\Delta_\pm = -\frac{3}{2}\pm \frac{1}{2}\sqrt{9+16\frac{f_1^2+16\cos^6\psi_\infty-12\cos^4\psi_\infty}{f_1^2+4\cos^6\psi_\infty}} \ .
\ee
Stability requires that $\Delta$ be real.  Demanding the square root be non-negative gives the following conditions:
\bea
f_1^2 & \geq & \frac{4}{25} \left(48\cos^4\psi_\infty -73 \cos^6\psi_\infty\right) \\
f_2^2 & \geq & \frac{4}{25} \left(48\sin^4\psi_\infty -73 \sin^6\psi_\infty\right) \,.
\eea

As explained in Section \ref{sec:setup} for the bulk system to have the correct holographic dual, the solution for $\psi$ must have an expansion for large $r$ of the form
\be
 \psi \sim \psi_{\infty}+\frac{m}{r}-\frac{c_{\psi}}{r^2} \ .
\ee
In other words, we require $\Delta_+ = -1$ and $\Delta_- = -2$, which means
\bea
\label{eq:f1squared_2}
 f_1^2 & = & 8\cos^4\psi_\infty -  12\cos^6\psi_\infty \\
\label{eq:f2squared_2}
 f_2^2 & = & 8\sin^4\psi_\infty -  12\sin^6\psi_\infty \,.
\eea

\section{Near-horizon asymptotics for BH embeddings}\label{app:nearhorizon}

To find BH solutions, we fix the initial $\psi_T \equiv \psi(r_T)$ and integrate the $\psi$ equation of 
motion (\ref{psieom3}) out from the horizon.  As (\ref{psieom3}) is second-order, 
two boundary conditions are required.  
However, the boundary condition for $\psi'_T$ is dictated by (\ref{psieom3}) since it becomes first-order 
in the near-horizon limit.

As we take $r \to r_T$ and $h \to 0$, equations (\ref{geqn}), (\ref{zeqn}), (\ref{Aeqn}) become
\bea
 g   & \to & \frac{h}{1+\frac{b^2}{r_T^4}}\sqrt{\tilde d_T^2+\left(r_T^4+b^2\right)\left(f_1^2+4\cos^4\psi_T\right)\left(f_2^2+4\sin^4\psi_T\right) } \\
 z'   & \to & \frac{f_1 f_2}{\sqrt{\tilde d_T^2 + \left(r_T^4+b^2\right)\left(f_1^2+4\cos^4\psi_T\right)\left(f_2^2+4\sin^4\psi_T\right)}} \\
 a'_0 & \to & \frac{\tilde d_T}{\sqrt{\tilde d_T^2 + \left(r_T^4+b^2\right)\left(f_1^2+4\cos^4\psi_T\right)\left(f_2^2+4\sin^4\psi_T\right)}} ,
\eea
where the horizon charge $\tilde d_T = d - 2 b \left(\psi_T - \psi_\infty - \frac{1}{4} \sin{4\psi_T} + \frac{1}{4} \sin{4\psi_\infty} \right)$.
The $\psi$ equation of motion (\ref{psieom3}) reduces to 
\bea 
  & & \left(\tilde d_T^2+r_T^4\left(1+\frac{b^2}{r_T^4}\right) \left(f_1^2+4\cos^4\psi_T\right) \left(f_2^2+4\sin^4\psi_T\right)\right)r_T \psi'_T = \nonumber \\
  & & -4\cos^2\psi_T\sin^2\psi_T b \tilde d_T \nonumber\\
  & & +2r_T^4 \left(1+\frac{b^2}{r_T^4}\right)\cos\psi_T\sin\psi_T\Big(f_1^2\sin^2\psi_T-f_2^2\cos^2\psi_T \nonumber\\
  & & \qquad\qquad\qquad\qquad\qquad\qquad+4\cos^2\psi_T\sin^2\psi_T\left(\cos^2\psi_T-\sin^2\psi_T\right)\Big) \ .
\eea
which gives the boundary condition for $\psi'_T$ at the horizon.

\section{Conductivity}
\label{app:conductivity}

In this appendix, we begin the computation of the electrical conductivity, performing the steps which are common to both the MN and BH embeddings.  In Section \ref{sec:BHconductivity}, we will directly apply the Karch-O'Bannon method \cite{Karch:2007pd, O'Bannon:2007in} (see also \cite{Lifschytz:2009si})
to the BH embeddings, while in Section \ref{MNconductivity}, we will use a generalization of that method for the MN embeddings.

While there is no true electromagnetic field on the field theory side, we employ the standard technique of modeling a background electric field holographically with a constant spacetime D7-brane gauge field.  The resulting charged currents are encoded in the gauge field's radial components.  Therefore, in addition to $f_1$, $f_2$, $b$, and $a'_0$, let us also turn on the following gauge fields:
\be
 F_{0x}= \frac{L^2}{2\pi\alpha'}e \quad , \quad F_{rx}= \frac{L^2}{2\pi\alpha'} a'_x(r),\quad , \quad F_{ry}= \frac{L^2}{2\pi\alpha'} a'_y(r) \ .
\ee
In this case, we have the DBI action, generalized from (\ref{DBI3}),
\be
 S_{DBI} = - \mathcal N \int d^3x dr \ r^2 \sqrt{(f_1^2+4\cos^4\psi)(f_2^2+4\sin^4\psi) Y} \ , 
\ee
where we have defined 
\bea
 Y & = & \left(1+\frac{b^2}{r^4}-\frac{e^2}{hr^4}\right)\left(1+hr^4z'^2+hr^2\psi'^2\right) \nonumber\\
   &   & -\left(1+\frac{b^2}{r^4}\right)a_0'^2+ha_x'^2+\left(1-\frac{e^2}{hr^4}\right)ha_y'^2-\frac{2eb}{r^4}a'_0a'_y \ .
\eea
For the CS action, in addition to the two terms we had in (\ref{CS2}), we also have
\bea
 (2\pi\alpha')^2C_{\theta_1\phi_1\theta_2\phi_2}F_{0x}F_{ry} &=& -\frac{L^8}{2} c(r) ea_y'  \sin\theta_1\sin\theta_2   \ .
\eea
The  CS action is now
\be
 S_{CS} = -\mathcal N \int d^3xdr \left(r^4 f_1 f_2 z'-2 c(r)\left(ba_0'+ea_y'\right) \right) \ .
\ee

From this action, we derive the equations of motion for the gauge fields.  First, generalizing (\ref{gdef3}) we define 
\be
\label{gdefconduct}
 g = hr^2 \sqrt{\frac{(f_1^2+4\cos^4\psi)(f_2^2+4\sin^4\psi)}{Y}} \ .
\ee
Since the action is independent of $a_\mu$, we can integrate the corresponding equations of motion to obtain
\bea
 \tilde d \equiv d-2b c(r) &=& \frac{g}{h}\left(\left(1+ \frac{b^2}{r^2}\right)a'_0+ \frac{eb}{r^4}a'_y\right) \\
 \label{axconduct}
 j_x&=& g a_x' \\
 \tilde j_y \equiv j_y-2c(r) e &=& \frac{g}{h}\left(\frac{eb}{r^4}a'_0-\left(1-\frac{e^2}{hr^4}\right) h a_y' \right) \ ,
\eea
and the integration constants $d$, $j_x$ and $j_y$ are conserved.  

We start by reshuffling gauge field equations to solve for $a_0'$ and $a_y'$:
\be
\label{a0ayconduct}
 a'_0=\frac{h\tilde d(1-\frac{e^2}{hr^4})+ \tilde j_y \frac{eb}{r^4}}{g(1+\frac{b^2}{r^4}-\frac{e^2}{hr^4})}\quad , \quad a'_y=\frac{\tilde d\frac{eb}{r^4}- \tilde j_y (1+\frac{b^2}{r^4})}{g(1-\frac{e^2}{hr^4}+\frac{b^2}{r^4})} \ .
\ee
Then, we use the equations for the gauge fields (\ref{axconduct}) and (\ref{a0ayconduct}) to solve (\ref{gdefconduct}) for $g$, obtaining 
\be
 g = \frac{\sqrt X}{(1-\frac{e^2}{hr^4}+\frac{b^2}{r^4})\sqrt{1+\frac{hr^4}{L^4}z'^2+hr^2\psi'^2}} \ ,
\ee
where
\bea
X   & = & h\left(1+\frac{b^2}{r^4}-\frac{e^2}{hr^4}\right)\left(hr^4(f_1^2+4\cos^4\psi)(f_2^2+4\sin^4\psi)+h\tilde d^2 - \tilde j_{y}^{2} - j_x^2\right) \nonumber \\ 
   &   & -(\sqrt h b \tilde d-e \tilde j_y)^2 \ .
\eea
When taking the square root to find $g$, we choose the plus sign, in accord with the original definition of $g$ being positive.  
Using this expression for $g$, we can write the gauge fields as
\bea
\label{A0appC_2}
 a'_0 & = & \left(h \left(1-\frac{e^2}{hr^4} \right) \tilde d+  \frac{eb}{r^4} \tilde j_y\right)\sqrt\frac{1+hr^4z'^2+hr^2\psi'^2}{X} \\
\label{A1appC_2}
 a'_x & = & j_x\left(1-\frac{e^2}{hr^4}+\frac{b^2}{r^4}\right)\sqrt\frac{1+hr^4z'^2+hr^2\psi'^2}{X} \\ 
 \label{A2appC_2}
 a'_y & = & \left(\frac{eb}{r^4} \tilde d-\left(1+\frac{b^2}{r^4}\right)\tilde j_y\right)\sqrt\frac{1+hr^4z'^2+hr^2\psi'^2}{X} \ .
\eea
At this point the MN and BH calculations diverge and are completed in their respective Sections.

\end{document}